\documentclass[11pt,a4paper]{article}

\usepackage{amsmath}
\usepackage{amssymb}
\usepackage{verbatim}
\usepackage[dvips]{graphicx}

\newcommand{\la}[1]{\label{#1}}
\newcommand{\be}{\begin{equation}}
\newcommand{\ee}{\end{equation}}
\newcommand{\ba}{\begin{eqnarray}}
\newcommand{\ea}{\end{eqnarray}}
\newcommand{\bi}{\begin{itemize}}
\newcommand{\ei}{\end{itemize}}
\newcommand{\bc}{\begin{center}}
\newcommand{\ec}{\end{center}}
\newcommand{\bmin}{\begin{minipage}}
\newcommand{\emin}{\end{minipage}}
\newcommand{\rmi}[1]{{\mbox{\scriptsize #1}}}
\newcommand{\fig}{Fig.~}
\newcommand{\eq}{Eq.~}
\newcommand{\se}{Sec.~}
\newcommand{\eqs}{Eqs.~}
\newcommand{\nr}[1]{(\ref{#1})}

\newcommand{\fr}[2]{{\frac{#1}{#2}\,}}

\newcommand{\bfx}{{\bf x}}

\newcommand{\qbar}{$\bar{\rm q}\;$}

\newcommand{\qs}{Q_{\mathrm{s}}}
\newcommand{\as}{\alpha_{\mathrm{s}}}
\newcommand{\gt}{{\gamma^0}}
\newcommand{\gz}{{\gamma^3}}
\newcommand{\half}{\frac{1}{2}}
\newcommand{\dtau}{{\partial_{\tau}}}
\newcommand{\deta}{{\partial_{\eta}}}
\newcommand{\ud}{\mathrm{d}}
\newcommand{\xt}{\mathbf{x}_T}
\newcommand{\yt}{\mathbf{y}_T}

\newcommand{\pt}{{\mathbf{p}_T}}
\newcommand{\qt}{{\mathbf{q}_T}}
\newcommand{\kt}{{\mathbf{k}_T}}

\newcommand{\Dt}{{\mathbf{D}_T}}

\newcommand{\nabt}{{\boldsymbol{\nabla}_T}}
\newcommand{\At}{{\mathbf{A}_T}} 
\newcommand{\gtrans}{\boldsymbol{\gamma}_T}
\newcommand{\Asl}{A \!\!\! /} 
\newcommand{\shat}{\hat{s}}

\begin{document}

\begin{titlepage}
\begin{flushright}
HIP-2004-38/TH\\
SPhT-T04/111\\
hep-ph/0409058\\
\end{flushright}
\begin{centering}
\vfill

{\Large{\bf Quark-antiquark production from classical fields in
heavy ion collisions: 1+1 dimensions}}

\vspace{0.8cm}

F. Gelis$^{\rm a}$            \footnote{fgelis@cea.fr},
K. Kajantie$^{\rm b}$         \footnote{keijo.kajantie@helsinki.fi},
T. Lappi$^{\rm b,c}$          \footnote{tuomas.lappi@helsinki.fi}.

\vspace{0.8cm}

{\em $^{\rm a}$%
  Service de Physique Theorique, B\^at. 774, CEA/DSM/Saclay, 91191
  Gif-sur-Yvette, France\\}

{\em $^{\rm b}$%
  Department of Physics, P.O.Box 64, FIN-00014 University of Helsinki,
  Finland\\}

{\em $^{\rm c}$%
  Helsinki Institute of Physics, P.O.Box 64, FIN-00014 University of Helsinki,
  Finland\\}

\vspace*{0.8cm}

\end{centering}

\noindent
Classical color fields produced by the small-$x$ wave functions of
colliding ultrarelativistic nuclei have been numerically computed.
We set up the framework for computing the production of small-mass
quark-antiquark pairs in these color fields by numerically integrating
the Dirac equation. This computation is essential for
understanding the conversion of the initial gluonic state to
chemically equilibrated quark-gluon plasma.  To illustrate and
overcome technical difficulties associated with the longitudinal
dimension, we first consider numerically the case of one time + one
longitudinal space dimension.
\vfill
\noindent

\vspace*{1cm}

\noindent

\vfill

\end{titlepage}

%
\section{Introduction}
The dynamics of an ultrarelativistic heavy ion collisions is usually
described in the following terms: two nuclei in their $T=0$, entropy=0
ground state move along the light cone, collide at $t=0$ and form a
large-entropy extended system with deconfined quark-gluon degrees of
freedom. This system expands, passes a QCD phase transition, converts
itself to a hadronic phase, which finally decouples and sends hadrons
to detectors.

Important partial confirmation for this scenario comes from recent
experimental results from the relativistic heavy ion collider RHIC.
These suggest that in $\sqrt{s}=$ 200 GeV Au+Au collisions a nearly
thermalised quark-gluon plasma is formed \cite{qm04}.  One of the main
pieces of evidence comes from azimuthal asymmetries in non-central
collisions \cite{starv2,phenixv2,phobosv2}: a hydrodynamic computation
\cite{kolbsollfrankheinz}, with an assumed equation of state and
initial conditions fitted to transverse spectra, shows that the
initial spatial azimuthal asymmetry is converted to just the correct
amount of momentum space azimuthal asymmetry if the equation of state
is that of an ideal fluid.

There is one significant deficiency in the theoretical analysis of
this scenario: virtually all the models describing the initial state
(for examples, see \cite{glr,kmw,mv1,ekrt,gm}) are based on the almost
purely gluonic small-$x$ partonic content of the nuclear wave
function, while an ideal quark-gluon plasma would contain gluons and
quarks+antiquarks in the ratio 16/(21$N_f$/2) and anyway the final
hadronic state contains flavour in a fully thermalised manner
\cite{braun-munzinger}. At what stage do the small-mass u,d, and s
flavour degrees of freedom appear in the system? Experiments do not
yet shed any light on this problem.

Models based on weakly coupled quark-gluon degrees of freedom, like
parton cascade models, fail to reproduce both kinetic and chemical
equilibration \cite{elliottrischke}; the coupling is so weak that
collision times become too large relative to the lifetime of the
system. The purpose of this paper is to start from a strongly coupled
and phenomenologically viable model, the classical field computation
of gluon production in a collision of two nuclei
\cite{kmw,krasnitzvenu,lappi} using the McLerran-Venugopalan model
\cite{mv1} for the distribution of gluons in a single nucleus,
later evolved and termed color glass condensate (see \cite{cgc} and references
therein).  We then discuss how
the amount of small-mass u,d, and s quark-antiquark pairs produced by
the colour fields of this model can be computed by numerically
integrating the evolution of a negative energy spinor as given by the
Dirac equation and projecting on a positive energy
spinor\cite{baltzmcl,baltz}. In this paper we give numerical results
only for a 1+1 dimensional toy model version of the full computation,
to establish the viability of the method.

The following should be emphasised from the outset:
\bi
\item This is not a computation of pair production from strong colour
  fields by quantum tunneling via the Schwinger mechanism, which has
  often been studied \cite{matsui,mottola}. Instead, the pairs are
  produced via multiple interactions of quasi-real Fourier components
  of the color fields; in the dilute limit this is just the two gluon
  fusion mechanism g$^\star$+g$^\star$$\to$ q+\qbar, which for heavy quarks
  also is the dominant mechanism \cite{gelisvenu}.  The same	
  production mechanism has been studied in \cite{dietrich}, where
  several approximations for the quark retarded propagator in an
  external field have been investigated.
\item Basically, there are two quantities we would like to know: how
  fast does the q\qbar density grow in comparison with the gluon
  density (what are the typical production times in units of $1/Q_s$, $Q_s$ =
  saturation scale, see below) and how high is the q\qbar density in
  comparison with the gluon density (what are the total energies per
  unit rapidity in units of $R_A^2Q_s^3$).  Parametrically, quark pair
  production is suppressed by a factor $\as$, but we are not in the
  weak coupling limit. 
  In the strong
  coupling regime, a large q\qbar component could be created, as
  required for chemical equilibration.
  Kinetic equilibration of the longitudinal degree of freedom
  is still an open issue.
\item The pair production being computed in a given colour field, the
  feedback is not taken into account.  The results will thus be
  quantitatively reliable only as long as the energy in q\qbar pairs
  remains less than that in gluons.  For the Abelian Higgs model in
  1+1 dimensions and with different initial conditions a numerical
  scheme for including both bosonic and fermionic dynamical degrees of
  freedom has been developed in \cite{aartssmit}.
\item The computation of the q\qbar production is technically much
  more complicated than that of gluons. The colour fields giving rise
  to gluons are assumed to be independent of the space-time rapidity
  $\eta=\fr12\log(x^+/x^-)$, they only depend on $\tau, \bfx_T$. Thus
  strict boost invariance for gluons is obtained (unless rapidity
  dependence is introduced via that of the saturation scales) and 
  the single rapidity of the problem, that of the gluon, can be
 completely removed from the equations.
  For quark pair
  production, two rapidities enter and a non-trivial dependence in
  $\Delta y=y_\mathrm{q}-y_{\bar{\mathrm{q}}}$ appears. This also
  implies that the quark wave function $\psi(\tau,\eta,\xt)$ will
  depend on all the 3+1 variables. However, formulating the initial
  condition on the light cones (say, $x^-=0, x^+>0$) is impossible
  using the natural variables $\tau,\eta$ since fixed
  $x^+=\tau\exp(\eta)/\sqrt2>0$ cannot be reached for $\tau\to0$
  unless also $\eta\to\infty$. We thus have to use as variables
  $\tau,x^\pm$ or, more symmetrically, $\tau,z$.

\item For an approach to quark pair production via special
  nonperturbative instanton configurations, see \cite{shuryakzahed}.

\ei

The structure of this paper is as follows. In section \ref{general}
the problem is formulated in full generality in 3+1 dimensions.
Particular attention is given to the initial condition and the
difficulties associated with the longitudinal dimension are pointed
out. This leads us to truncate the full theory by neglecting all the
transverse integrations to a 1+1 dimensional toy model, with which we
can test the numerical solution of the time+longitudinal dependence.
The free Dirac equation in 1+1 dimensions using ($\tau,\eta$),
($\tau,x^\pm$) or ($\tau,z$) as variables (we do not go all the way to
($t,z$)!) is studied and solved analytically in section
\ref{free1+1analytic}. Its numerical solution is carried out in
section \ref{free1+1numerical} and shown to agree with the analytic
one. Finally, in section \ref{1+1external} the 1+1 dimensional Dirac
equation is solved with various forms of the external gluonic field.

In the time-longitudinal space we shall use three sets of variables:
$t,z$, $\ud s^2= \ud t^2- \ud z^2$, the light cone coordinates $x^\pm
= (t \pm z)/\sqrt{2}=\tau e^{\pm\eta}/\sqrt2$, $\ud s^2=2\ud x^+\ud
x^-$ and proper time and spacetime rapidity $\tau = \sqrt{t^2-z^2}=
\sqrt{2x^+ x^-}$, $\eta = \half \ln (x^+/x^-)$, $\ud s^2 = \ud \tau^2
- \tau^2 \ud \eta^2$. For any four-vector $A_\mu$ we have as $A_\tau
=A^\tau= (t A^0 - z A^3 )/\tau=(x^+A^-+x^-A^+)/\tau$ and $A_\eta =
-\tau^2 A^\eta = z A^0 - t A^3=x^+A^--x^-A^+$.  This also applies to
Dirac gamma matrices, giving $\gamma^\tau = \gt e^{-\eta \gt \gz}.$ We
shall frequently separate Dirac spinors into eigenvectors of $\gt
\gz,$ using the projection operators 
\be 
P^\pm = \half (1 \pm \gt
\gz)=\frac{1}{\sqrt2}\gt \gamma^\pm= \frac{1}{\sqrt2}
\gamma^\mp\gt=\fr12\gamma^\mp\gamma^\pm,
\label{P}
\ee satisfying $P^\pm P^\pm = P^\pm,$ $P^\pm P^\mp =0$, $P^++P^-=1$.
For momenta we use the transverse mass $\omega_p \equiv \pt^2 + m^2$
and rapidity $y = \half \ln (p^+/p^-),$ giving the energy $E_p =
\omega_p \cosh y$ and the longitudinal momentum $p^z = \omega_p \sinh
y.$

\section{General formulation of the problem in 3+1 dimensions; initial
conditions on the light cone}
\la{general}

\begin{figure}[!htb]
\begin{center}
\includegraphics[width=0.49\textwidth]{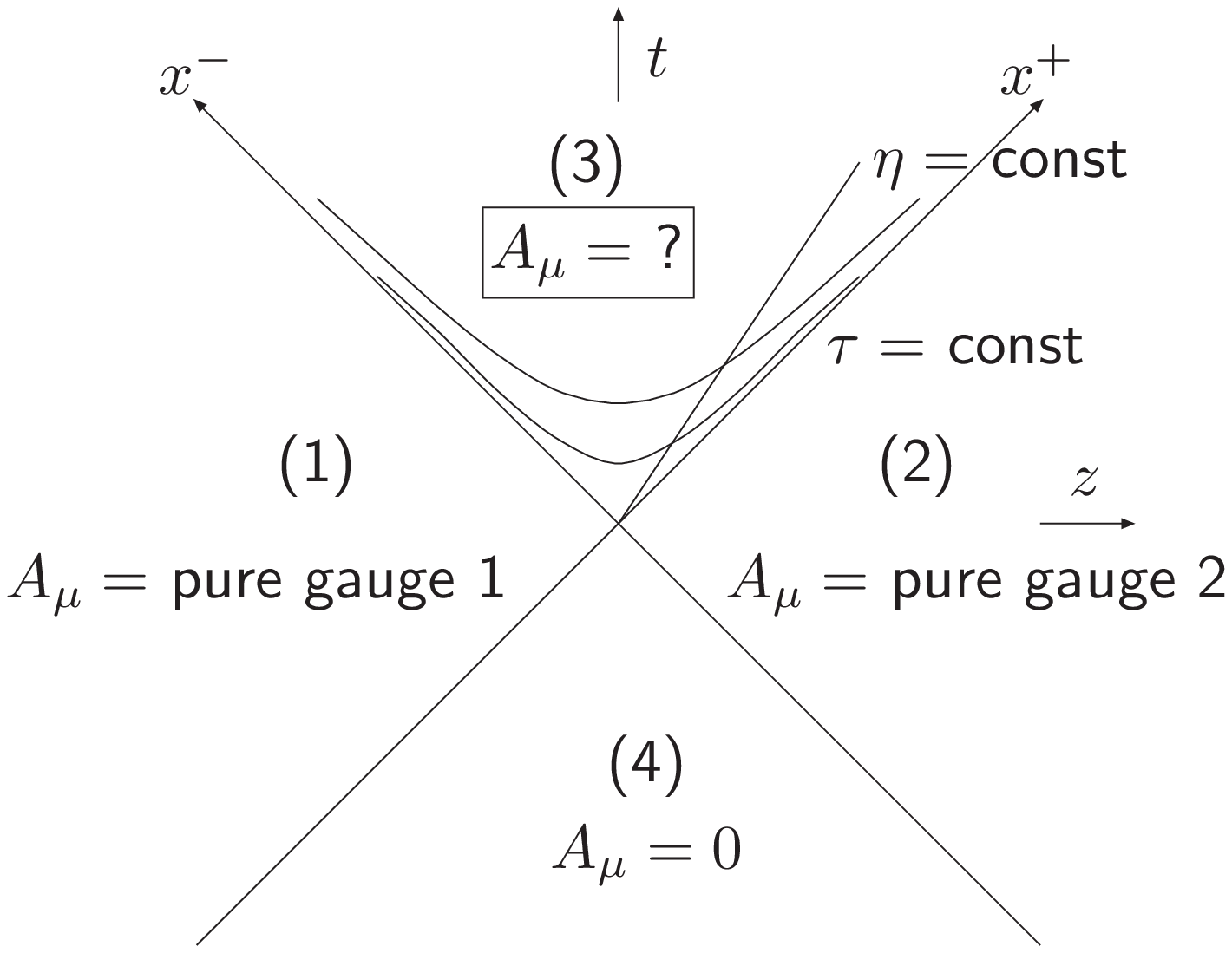}
\includegraphics[width=0.49\textwidth]{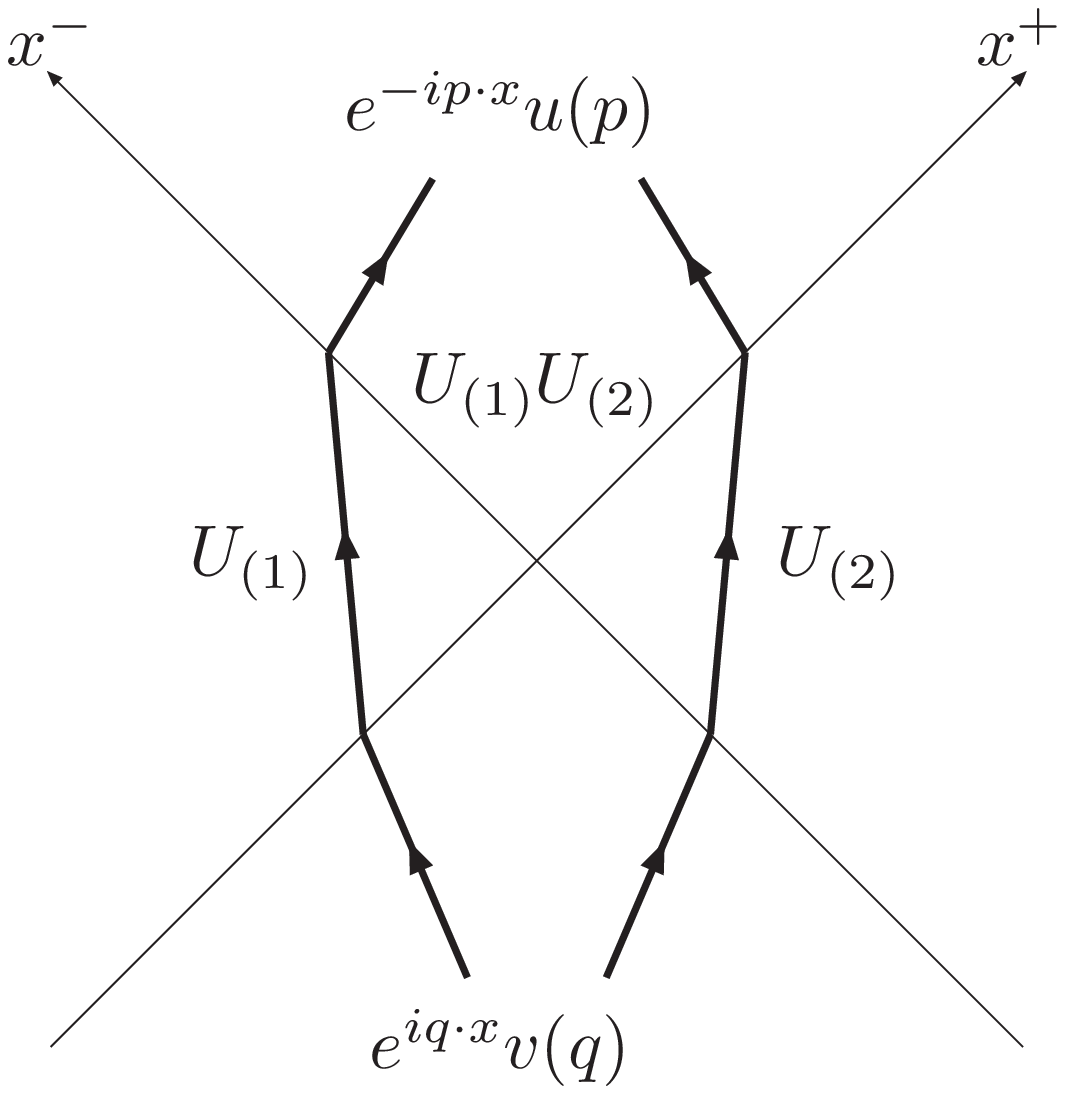}
\end{center}
\caption{\small Left: The background gauge field from 
  the classical field model.  In the Abelian case the field is a pure
  gauge also in region (3).  Right: The two possible trajectories of
  (retarded) propagation of the fermion.  The spinor starts from a
  negative energy state $e^{i q \cdot x}v(q)$ and gets a $k^+$-kick
  from the gluon field (1) on the $x^+$-axis and then a $k^-$-kick
  from the gluon field (2) on the $x^-$-axis; or in the other order.
  Finally the spinor is projected onto positive energy states $e^{-i p
    \cdot x}u(p)$ inside the future light cone. The label
  $U_{(1)}U_{(2)}$ refers only to the Abelian case, when the pure
  gauge field inside the future light cone is given by
  $U_{(1)}U_{(2)}= U_{(2)}U_{(1)}.$ }\label{fig:spacet}
\end{figure}

We want to calculate the number of quark-antiquark pairs produced by
the classical color fields in the model developed in
Refs.~\cite{kmw,mv1} and solved numerically in
Refs.~\cite{krasnitzvenu,lappi}.  In \cite{baltz}, it was shown that
the average number of produced pairs can be expressed as follows:
\begin{eqnarray}
\big<n_{q\bar{q}}\big>
&=&
\int\frac{d^3{\bf p}}{(2\pi)^3 2 E_p}
\Big<0_{\rm in}\Big|
b^\dagger_{\rm out}({\bf p})
b_{\rm out}({\bf p)}
\Big|0_{\rm in}\Big>
\nonumber\\
&=&
\int\frac{d^3{\bf p}}{(2\pi)^3 2 E_p}
\int\frac{d^3{\bf q}}{(2\pi)^3 2 E_q}
\Big|
\overline{u}(p)
{\cal T}_{_{R}}(p,-q)
v(q)
\Big|^2,
\end{eqnarray}
where ${\cal T}_{_{R}}(p,-q)$ is the amputated retarded propagator of
the quark in the external field, $-q$ and $p$ being respectively the
incoming and outgoing momenta. Note that this quantity is distinct
from the cross-section to produce one pair, which would require the
time-ordered propagator of the quark. The difference between the two
cases is explained in more detail in \cite{baltz}. For computational
purposes, it is in fact more convenient to write the previous formula
in coordinate space. This can be achieved by using the standard
machinery of reduction formulas, and one obtains:
\begin{equation}\label{eq:melement}
\overline{u}(p)
{\cal T}_{_{R}}(p,-q)
v(q)
=\lim_{x_0\to +\infty}
\int d^3{\bf x}
e^{i(E_p x^0-{\bf p}\cdot{\bf x})}u^\dagger(p)
\psi_{\bf q}(x^0,{\bf x}),
\end{equation}
where $\psi_{\bf q}(x^0,{\bf x})$ is the solution of the Dirac
equation in the presence of the external field, with a negative energy
free spinor as its initial boundary value:
\begin{equation}\label{eq:asymp}
\psi_{\bf q}(x^0\to-\infty,{\bf x})=
e^{i(E_q x^0-{\bf q}\cdot{\bf x})}v(q) .
\end{equation}
This formula can be trivially modified in order to use the proper time
$\tau$ instead of $x^0$. Moreover, although it tells us that the
on-shell pair production amplitude is obtained only in the limit of
infinite time, one can also consider an extension of this formula
where the limit is not taken, thereby defining a ``time dependent
pair-production amplitude'' which could be used in order to probe the
typical time-scale necessary to produce the fermions. We shall give
several examples of this below (see \fig\ref{fig:tausize} and
\fig\ref{fig:expdecay}).

Note that in this formalism, one neglects the backreaction of the
produced fermions on the color field. It is therefore a good
approximation only if the fermions do not outnumber the gluons.  In an
Abelian theory the calculation only involves pure gauge fields and the
pair production amplitude (see \eq\nr{eq:bmcl} below) can be
calculated analytically\footnote{However, the evaluation of this
  amplitude is complicated by some infrared singularities that need to
  be properly regularized \cite{leemilstein}.}, as shown in
\cite{baltzmcl}. Furthermore, in an Abelian theory, the square of this
generalized amplitude is, in fact, time-independent, suggesting that
all the pairs are produced instantaneously at the collision time (for
a collision at infinite energy).

In the non-Abelian case the gauge fields are
known analytically up to the future light cone ($\tau=0$) and
numerically for $\tau \geq 0.$ Thus we will have to solve the
Dirac equation numerically for that region, with the initial
condition at $\tau=0$ given by essentially the same calculation as
in the Abelian case.

Let us first review the classical field model of \cite{kmw,mv1}. Then we shall
repeat the calculation of fermion-antifermion production from Abelian
Weizs\"acker-Williams fields from Ref.~\cite{baltzmcl} and use it to
find the initial conditions at $\tau=0$ for our numerical calculation
in the non-Abelian case.

Let us assume we have two nuclei moving along the light cone, corresponding to a current
\be
J^\mu=\delta^{\mu+}\delta(x^-)\rho_{(1)}(\xt)+
\delta^{\mu-}\delta(x^+)\rho_{(2)}(\xt).
\la{current}
\ee
The two colour charge densities $\rho_{(m)}(\xt)$ are, independently for
the two nuclei, drawn from a random
ensemble, which in the original McLerran-Venugopalan model is taken to
be Gaussian:
\be
\langle \rho^a_{(m)}(\xt)\rho^b_{(m)}(\yt)\rangle=
g^2\mu_{(m)}^2\delta^{ab}\delta^2(\xt-\yt),
\quad m=1,2,
\label{rhorho}
\ee
where $\mu$ is a parameter describing the transverse density of color charges
and can be related, up to a logarithmic uncertainty, to the saturation
scale $\qs$ \cite{Krasnitz:2002mn}. More generally the charge distribution in
\eq\nr{rhorho} is not known, but its evolution when probing smaller Feynman $x$
values in the nuclei can be calculated from the JIMWLK equation, see
e.g. \cite{jimwlk}.

One first calculates in the light cone gauge 
($A^+=0$ for the nucleus moving in the $+z$ direction, and $A^-=0$ for the
nucleus moving in the $-z$ direction)
the pure gauge fields corresponding two the two nuclei:
\be \label{eq:pureg}
A^i_{(m)}(\xt)=\frac{i}{g}U_{(m)}(\xt)\partial_i U_{(m)}^\dag(\xt),\quad m=1,2.
\ee
These depend on the Wilson lines $U_{(m)}(\xt)$ given by
\be
U_{(m)}(\xt)=\exp \left(-ig \frac{\rho_{(m)}}{\nabt^2}(\xt)\right).
\ee
In a temporal gauge\footnote{The
$A_\tau=0$ gauge coincides with the light-cone gauge $A^+=0$ (resp. $A^-=0$)
if $x^+=0$ (resp. $x^-=0$). This is why we can use the gauge field of the
nuclei before the collision, in two different light-cone gauges, as an
initial condition at $\tau=0$ for the gauge field in the $A_\tau=0$ gauge.}
$A_\tau=0$ the initial condition at $\tau=0$
for the color fields  $\At(\tau,\xt)$ and $A_\eta(\tau,\xt)$ is given by these
pure gauge fields corresponding to the two nuclei:
\ba \label{eq:gaugeinit}
A^i(0,\xt)&=&A^i_{(1)}(\xt)+A^i_{(2)}(\xt),\nonumber\\
A^\eta(0,\xt)&=&\frac{ig}{2}[A^i_{(1)}(\xt),A^i_{(2)}(\xt)]. \ea
One then solves the equations of motion 
\be
[D_\mu,F^{\mu\nu}]=0
\la{eom} 
\ee 
using these initial
conditions to find the fields at later times $\tau>0.$ In this
gauge it is easy to find the Hamiltonian and thus the energy of a
given field configuration. Additionally, fixing the Coulomb gauge
in the transverse plane, $\nabt \cdot \At = 0,$ one can also
define a multiplicity corresponding to the classical fields.

Let us then turn to solving the Dirac equation in the Abelian case,
following \cite{baltzmcl}.  In \cite{baltzmcl} it is solved separately
in different gauges, covariant gauge (called singular gauge in
\cite{baltzmcl}) and light-cone gauge.  In \cite{gelisvenu} the covariant
gauge is used. Here we use the light-cone gauge, because it is the
same gauge that was used in solving the Yang-Mills equations up to the
$\tau=0$ light cone.

Following \eqs\nr{eq:melement} and \nr{eq:asymp}
we start with a negative energy plane wave for $x^\pm<0:$ 
\be
\psi_{(4)}(x) = e^{i q \cdot x } v(q). 
\ee 
The boundary conditions 
when crossing the light cones can be derived from the following argument.  Since
$\gamma^+P^-=\gamma^-P^+=0$, the Dirac equation only involves terms
like $\partial_-P^-\psi$ and $\partial_+P^+\psi,$ but not
$\partial_-P^+\psi$ and $\partial_+P^-\psi$. If there were a
discontinuity in $P^- \psi$ on the $x^-=0$ light cone, the derivative
term would give a delta function, with no other term to compensate it,
which is not possible. A discontinuity in $P^+\psi,$ on the other
hand, is possible because it can be compensated by the
$\theta(x^-)$-discontinuity in the gauge field.  Thus the boundary
condition is that on the $x^\pm=0$ light cone $\psi^\pm$ is
continuous. Using this boundary condition one can find the solutions
in the regions (1) ($x^->0>x^+$) and (2) ($x^+>0>x^-$):
\begin{eqnarray}\la{eq:beforezero}
&&\psi_{(1)}(x) = U_{(1)}(\xt) \int \frac{\ud^2\kt}{(2\pi)^2}U_{(1)}^\dag(\kt-\qt)
e^{-i \kt \cdot \xt}
\nonumber\\
&&\qquad\qquad\times
\exp \left( 
i q^- x^+
+i \frac{\omega^2_k}{2q^-} x^-\right)
\left[ P^- + P^+ \gt \frac{\gtrans\cdot \kt - m}{\sqrt{2}q^-} \right]v(q)\; ,
\nonumber\\
&&
\psi_{(2)}(x)  = U_{(2)}(\xt) \int \frac{\ud^2\kt}{(2\pi)^2} U_{(2)}^\dag(\kt-\qt)
e^{-i \kt \cdot \xt}
\nonumber\\
&&\qquad\qquad\times\exp \left( 
i \frac{\omega^2_k}{2q^+} x^+
+i q^+ x^-
\right)
\left[ P^+ + P^- \gt \frac{\gtrans\cdot \kt - m}{\sqrt{2}q^+} \right] v(q)\; ,
\end{eqnarray}
where $\omega_k^2\equiv \kt^2+m^2$ and we have defined the Fourier
transforms as: 
\be \label{eq:four} U^\dag_{(m)}(\kt) \equiv \int \ud
^2 \yt e^{i \yt \cdot \kt}U_{(m)}^\dag(\yt).  
\ee 

Next we must continue to
the forward light cone. The solution that matches to
\eq\nr{eq:beforezero} on the light cone is
\begin{eqnarray}\label{eq:fullsol}
\psi_{(3)}(x) &=&
U_{(1)}(\xt) U_{(2)}(\xt) \int \frac{\ud^2\pt}{(2\pi)^2}
\frac{\ud^2\kt}{(2\pi)^2}
\left[ \frac{\ud p^+}{2\pi i} \frac{1}{p^+ - \frac{\omega^2_k}{2q^-} - i \epsilon}\right]e^{-i\pt\cdot\xt}
\nonumber\\
&&\quad\times
\exp \left(
i \frac{\omega^2_p}{2p^+ - i\epsilon} x^+
+ i p^+ x^- \right)
U_{(2)}^\dag(\pt-\kt)U_{(1)}^\dag(\kt -\qt)
\nonumber\\
&&\quad\times
\left[ P^+ + P^- \gt \frac{\gtrans\cdot \pt - m}{\sqrt{2}p^+ - i \epsilon} \right]
\gt \frac{\gtrans\cdot \kt - m}{\sqrt{2}q^-} v(q)
\nonumber\\
&&
+U_{(1)}(\xt) U_{(2)}(\xt) \int \frac{\ud^2\pt}{(2\pi)^2}
\frac{\ud^2\kt}{(2\pi)^2}
\left[ \frac{\ud p^-}{2\pi i} 
\frac{1}{p^- - \frac{\omega^2_k}{2q^+} - i \epsilon}\right]
e^{-i\pt\cdot\xt}
\nonumber\\
&&\quad\times
\exp \left(
i \frac{\omega^2_p}{2p^- - i \epsilon} x^-
+i p^- x^+
\right)
U_{(1)}^\dag(\pt-\kt) U_{(2)}^\dag(\kt-\qt)
\nonumber\\
&&\quad\times
\left[ P^- + P^+ \gt \frac{\gtrans\cdot \pt - m}{\sqrt{2}p^- - i \epsilon} \right]
\gt \frac{\gtrans\cdot \kt - m}{\sqrt{2}q^+} v(q)\; .
\end{eqnarray}
Here the first term corresponds to a situation in which the positron
state $q$ first hits the nucleus 1 moving in the $x^+$ direction,
propagates over region 1, meets the nucleus 2 and propagates into
region 3 (the branch on the left in \fig\ref{fig:spacet}).  For QED
the order of the $U_{(m)}$ matrices is irrelevant, not so for QCD. The
$p^\pm$-integrals in \eq\nr{eq:fullsol} can be performed to turn the
expression into a sum of Bessel functions of the kind we shall
encounter in Sec.~\ref{free1+1analytic}, but we will not write down
this complicated expression here.

To find the matrix element for pair production, one has to project the
spinor \nr{eq:fullsol} to a positive energy state 
$e^{-ip \cdot x}u(p)$.  Removing the product $U_{(2)}U_{(1)}$, which is a gauge
transformation to Coulomb gauge, one gets the Abelian theory result of
\cite{baltzmcl}:
\begin{multline}\label{eq:bmcl}
M(p,q) =
i \sqrt{2} \int \frac{\ud^2\kt}{(2\pi)^2}
\Bigg\{
\frac{U_{(2)}^\dag(-\pt - \kt)U_{(1)}^\dag(\kt - \qt)}
{\omega_q \omega_p e^{y_p - y_q} +\omega_k^2 }  u^\dag (p)  \gamma^-
\left(\gtrans\cdot \kt - m\right)v(q)
\\
+
\frac{U_{(1)}^\dag(- \pt  - \kt)U_{(2)}^\dag(\kt - \qt)}
{\omega_q \omega_p e^{y_q - y_p} +\omega_k^2}  u^\dag (p)  \gamma^+
\left(\gtrans\cdot \kt - m\right)v(q)
\Bigg\}.
\end{multline}

\begin{figure}[!ht]
\begin{center}
\includegraphics[width=0.7\textwidth]{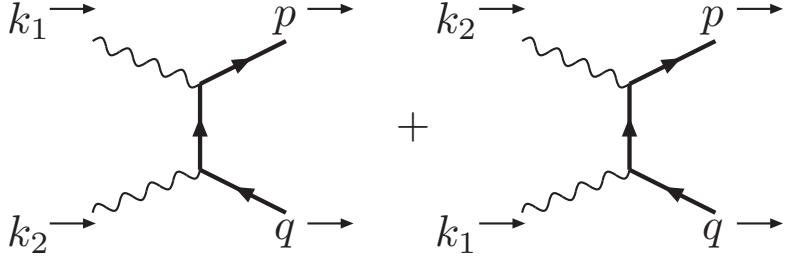}
\end{center}
\caption{\small Diagrams contributing to the lowest order pair 
  production amplitude in QED.  The incoming photons are quasi-real,
  with $k_1=(p^+ + q^+,0,\kt+\pt)$ and $k_2=(0,p^- + q^-,\qt-\kt).$ }
\label{fig:tchan}
\end{figure}

Kinematically, the terms in \nr{eq:bmcl} correspond to the process
$k_1+k_2\to p+q$ with $k_1=(p^+ + q^+,0,\kt+\pt)$, $k_2=(0,p^- +
q^-,\qt-\kt)$ (see \fig\ref{fig:tchan}). The two terms are the
$t$- and $u$-channel propagator pole terms in the Feynman diagram
corresponding to \fig\ref{fig:tchan}, with $m^2-t =
m^2-(k_1-p)^2=\omega_q\omega_p e^{y_q-y_p}+\omega_k^2$ and $m^2-u
= m^2-(k_1-q)^2=\omega_q\omega_p e^{y_p-y_q}+\omega_k^2.$

We now wish to take the Abelian solution for $\tau>0$,
\eq\nr{eq:fullsol}, and write it in a form suitable for determining
the initial condition as $\tau \to 0.$ What is crucial here is the 
choice of the other variable, kept fixed when taking this limit. The
choices are $\eta,z,x^\pm$.  To obtain the correct result we must have
a dimensionful longitudinal variable, such as $z$ or $x^\pm$ to
parametrise the $\tau=0$ surface.  This is because one must be able to
represent longitudinal momentum scales, for example $\omega_k^2
e^{y_q}/ \omega_q$, in coordinate space. For $\tau > 0$ the
corresponding longitudinal coordinate could be constructed as $\tau
e^{-\eta},$ but for $\tau=0$ this is not possible. To enable a
symmetric treatment of both branches in \fig\ref{fig:spacet}, we
choose $z$ as the longitudinal variable, with $x^\pm =
(\sqrt{\tau^2+z^2}\pm z)/\sqrt2=(|z|\pm z)/\sqrt{2}$ at $\tau=0.$
After a rather lengthy computation, the $\tau\to0$ limit of the wave
function \eq\nr{eq:fullsol} can be written as
\begin{eqnarray}\label{eq:allinitcond}
&&\psi_{(3)}(\tau=0, z,\xt) =
e^{-i \qt \cdot \xt} 
\int \frac{\ud^2\kt}{(2\pi)^2}
\;\;e^{-i \kt \cdot \xt}
\nonumber\\
&&
\times\Bigg\{ 
 P^+  \frac{e^{y_q}}{\omega_q}  U_{(1)}(\xt)
U_{(1)}^\dag(\kt)
\exp \left(i\frac{\omega_{k+q}^2 e^{y_q} (|z|-z)} {2 \omega_q} \right)
\nonumber\\ 
&&
\quad{+}
P^- \gt \left[i \gtrans \cdot \Dt -m \right]
U_{(1)}(\xt) U_{(1)}^\dag(\kt)
\frac{1}{\omega_{k+q}^2}
\left[ \exp \left(i\frac{\omega_{k+q}^2 e^{y_q} (|z|-z)} {2\omega_q} \right)
-1 \right]
\nonumber\\ 
&&
\quad{+}
 P^-  \frac{e^{-y_q}}{\omega_q} U_{(2)}(\xt)
U_{(2)}^\dag(\kt)
\exp \left(i\frac{\omega_{k+q}^2 e^{-y_q} (|z|+z)} {2 \omega_q} \right)
\nonumber\\ 
&&
\quad{+}
P^+ \gt \left[i \gtrans \cdot \Dt -m \right]
U_{(2)}(\xt)
U_{(2)}^\dag(\kt)
\frac{1}{\omega_{k+q}^2}
\left[ \exp \left(i\frac{\omega_{k+q}^2 e^{-y_q} (|z|+z)} {2 \omega_q}\right)
-1 \right]
\Bigg\}
\nonumber\\
&&
\qquad\times
\gt \left(\gtrans\cdot (\kt+\qt) - m\right)v(q),
\end{eqnarray}
where $\Dt=\nabt+ig{\mathbf A_T(0,\xt)}$ and
$\omega_{k+q}^2=(\kt+\qt)^2+m^2$.  Note that whereas
\eq\nr{eq:fullsol} involved products of $U_{(1)}$ and $U_{(2)}$,
implicitly assuming that they commute, the terms in
\eq\nr{eq:allinitcond} only contain either $U_{(1)}$ or $U_{(2)}$ and
thus \eq\nr{eq:allinitcond} can be directly generalised to the
non-Abelian theory. The products of $U_{(1)}$ and $U_{(2)}$ show up in
the gauge field in the covariant derivative $\Dt$, which depends on
both $U_{(1)}$ and $U_{(2)}$ by \eqs\nr{eq:pureg},\nr{eq:gaugeinit}.

Now we have the initial conditions for a numerical solution of the
Dirac equation in 3+1 dimensions. To proceed further, one will have to
generate the random SU(3) matrices $U_{(m)}(\xt)$, compute the color
fields $A_\eta(\tau,\xt), A_i(\tau,\xt)$, compute the spinor
$\psi_{(3)}(\tau,z,\xt)$ from the Dirac equation using these color
fields and the initial condition \eq\nr{eq:allinitcond}, project to a
positive energy state $e^{-ip\cdot x}u(p)$ and, finally, integrate the
square of the amplitude so obtained over momenta. As this is a rather
involved operation, we shall first simplify the problem by neglecting
the transverse dimension. Computation of q\qbar production in this 1+1
dimensional toy model permits us to test numerical aspects of the
solution and the projection to final states.  We shall return to the
3+1 dimensional case in future work.

As a first step, setting $U_{(m)}(\xt)=1,$ $U_{(m)}(\kt)= (2\pi)^2
\delta^2(\kt),$ in \eq\nr{eq:allinitcond} gives \ba \psi_{(3)}(\tau=0,
z,\xt)|_{U=1}=e^{-i\qt\cdot\xt} \biggl[ &&\hspace{-0.6cm}
e^{iq^+x^-}P^+-\frac{m}{\omega_q}e^{-y_q}\bigl(e^{iq^+x^-}-1\bigr)\gt
P^+ \biggr.
\nonumber\\
&&\hspace{-1cm}
\biggl.+e^{iq^-x^+}P^--\frac{m}{\omega_q}e^{y_q}\bigl(e^{iq^-x^+}-1\bigr)\gt
P^-\biggr]v(q)
\label{U=1}
\ea
This also illustrates the structure of \eq\nr{eq:allinitcond}: the left branch
(first line) has for $z<0$ (on the $x^-$-axis) firstly a component
$P^+v(q)$ moving in the $+z$ direction, but one
also needs a component $\sim \gt P^+v(q)$ to satisfy the Dirac equation.
This is worked out explicitly below in \eq\nr{eq:psimfree}. For $z>0$ (on the
$x^+$ axis) only $e^{-i\qt\cdot\xt}P^+v(q)$ moving in $+z$ direction remains.
The right branch behaves symmetrically.

\section{Free Dirac equation in 1+1 dimensions; analytic}
\la{free1+1analytic} Let us first define our spinor conventions and
study the solutions of the free Dirac equation in 1+1 dimensions.

In 1+1 dimensions, given fermions of mass $m$, we can parametrise an on-shell
momentum vector by just the rapidity $y$: $(E,p^z) = m (\cosh y ,
\sinh y).$ A free wave is then $e^{i p \cdot x } = e^{i m \tau \cosh
  (y-\eta) }.$

Let us choose for the Dirac matrices a representation where $\gt \gz$ is diagonal:
\begin{equation}
\gt = \sigma^1 = \left( \begin{array}{rr} 0 &  1 \\ 1  & 0 \end{array} \right)
, \quad
\gz = -i\sigma^2 = \left( \begin{array}{rr} 0 &  -1 \\ 1  & 0 \end{array} \right),
\quad
\gt \gz = \sigma^3 = \left( \begin{array}{rr} 1 &  0 \\ 0  & -1 \end{array} \right).
\end{equation}
In this basis the projection operators defined in \eq\nr{P} are simply:
\begin{equation}
P^+ = \left( \begin{array}{rr} 1 &  0 \\ 0  & 0 \end{array} \right),
\quad
P^- = \left( \begin{array}{rr} 0 &  0 \\ 0  & 1 \end{array} \right),
\end{equation}
and we denote the two components by $\psi^\pm$:
\be
\psi = \binom{\psi^+}{\psi^-}.
\ee

The Dirac equation:
\begin{equation}
\left(i \gamma^\mu \partial_\mu -m \right) \psi = 0
\end{equation}
has plane wave solutions corresponding to positive and negative energy:
\begin{equation}\la{eq:planew1}
\psi_{(+)} (x) = e^{-i p \cdot x} u(y)
\quad
\psi_{(-)} (x) = e^{i p \cdot x} v(y).
\end{equation}
Using the explicit forms of the Dirac matrices we can easily see that
\begin{equation}\la{eq:planew2}
u(y) = \sqrt{m} \binom{e^{\half y}}{e^{-\half y} },
\quad
v(y) = \sqrt{m} \binom{e^{\half y}}{- e^{-\half y} }.
\end{equation}
The solutions have a Lorentz-invariant normalisation with
$\overline{u}(y)u(y) = \overline{v}(y)v(y) =2m$ and
$\overline{u}(y)v(y) = \overline{v}(y)u(y) = 0.$ The quantity we will
be interested in, however, is the particle number density on a
constant proper time surface, which is the $\tau$--component of a
Lorentz vector. For both the positive and negative energy solutions it
is given by $\overline{u}(y) \gamma^\tau u(y) = \overline{v}(y)
\gamma^\tau v(y) = 2 m \cosh(y-\eta) $.  The cross term
$\overline{u}(y) \gamma^\tau v(y) = 2 m \sinh(y-\eta)$ is not zero,
but vanishes by symmetry when integrated over the longitudinal
coordinate $\eta$.

Writing the free Dirac equation in terms of 
$\psi^\pm = P^\pm \psi,$ the eigenvectors of $\gt \gz,$ we have
\begin{equation}\la{eq:freediracpm}
i \left( \dtau +  \frac{\deta}{\tau} \right)
\psi^\pm  =  m   e^{\pm\eta} \psi^\mp,
\end{equation}
or, squaring to get a second order equation,
\begin{equation}
\left[
\dtau^2 + \frac{1}{\tau}\dtau -
\frac{\deta^2}{\tau^2} + m^2 \right] \psi^\pm = 0.
\end{equation}
With an ansatz $\psi^\pm(\tau, \eta) = e^{n \eta} \psi^\pm_n(\tau)$
this reduces to the Bessel equation. The solutions that are separable
and finite for $\tau =0 $ are of the form $e^{n \eta}J_n(m \tau).$

To find the right linear combination of the separable solutions we
have to look at the initial condition. This can be done by looking at
the full initial condition, \eq\nr{eq:allinitcond}, and removing all
the transverse degrees of freedom.  One first takes $U_{(m)}(\xt)=1,$
which leads to \eq\nr{U=1}, and further sets $\qt=0$ and thus reduces
$\omega_q$ to $m$. Using the 2d representation of $\gt,\gz$ and $v(q)$
given above, \eq\nr{U=1} then becomes \footnote{We will
  actually change the sign of this expression; this would correspond
  to changing the sign of $v(q)$ in \eq\nr{eq:planew2}, which is just
  a convention.}
\begin{equation}
\psi(\tau=0, z) = 
\sqrt{m} 
\left(
\begin{array}{c}
 e^{y/2}  e^{i m e^{y} (|z|-z) /2 } \\
-e^{-y/2} ( e^{i m e^{y} (|z|-z) / {2} } -1  )
\end{array}
\right)
+ \sqrt{m} 
\left(
\begin{array}{c}
 e^{y/2} \left( e^{i m e^{-y} (|z|+z) / 2 } -1  \right) \\
  -e^{-y/2}  e^{i m e^{-y} (|z|+z) /2}
\end{array}
\right),
\label{1dinitcond}
\end{equation}
where the first(second) term is the left(right) branch in \fig\ref{fig:spacet}.
From this we find that the initial condition for, for example,
the upper component of the left hand branch,
must behave, up to a sign, for $\tau \to 0$ as:
\begin{equation}
\psi^+(\tau\to0,\eta) = - \sqrt{m} e^{y/2}  e^{i \half m \tau e^{y-\eta}}
= - \sqrt{m} e^{y/2}  e^{i \half m e^y (\sqrt{\tau^2+z^2}-z)}.
\la{tau0limit}
\end{equation}
Such a solution can be constructed as a sum of
Bessel function modes as
\begin{equation}\la{eq:psipfree}
\psi^+(\tau,\eta) = - \sqrt{m} e^{y/2} \sum_{n=0}^\infty (i e^{y-\eta})^n J_n(m\tau);
\end{equation}
using $J_n(m\tau)\to (m\tau/2)^n/n!$ for $\tau\to0$ and summing over
$n$ gives \eq\nr{tau0limit}.  Using \eq \nr{eq:freediracpm} we see
that the other component of the spinor is:
\begin{equation}\la{eq:psimfree}
\psi^- = \sqrt{m} e^{-y/2} \sum_{n=1}^\infty (i e^{y-\eta})^n J_n(m\tau),
\end{equation}
which is exactly the same as the lower component of the first term in
\eq\nr{1dinitcond}, showing the consistency of the approach.
We show in Appendix \ref{sec:ampli} that when this wave function is projected
on positive energy states, one obtains the same result ($\pm i/\cosh\fr12(y-y')$
for the two branches) for the amplitude
as from evaluating the Feynman diagrams in \fig\ref{fig:tchan} or by
specialising the Abelian amplitude in \eq\nr{eq:bmcl} to 2d.

\section{Free Dirac equation in 1+1 dimensions; numerical}
\la{free1+1numerical}

Let us now formulate the discretised solution of the Dirac equation
using the coordinates $\tau,z.$ The advantage of this coordinate
system is that it allows a simultaneous and symmetric treatment of
both two branches.  On the other hand, at $\tau=0,$ $\sqrt{2} x^\pm =
|z|\pm z $ is not a continuous function of $z$ and there are
corresponding discontinuities in $\psi$.  The free Dirac equation in
this coordinate system is
\begin{equation}\la{eq:eomtauz}
\dtau \psi^\pm = \frac{\sqrt{\tau^2+z^2}\pm z}{\tau}
\left( \mp \partial_z \psi^\pm
- im \psi^\mp \right)\; .
\end{equation}
The initial condition for the left hand branch is:
\begin{eqnarray}
\psi^+(\tau=0,z) &=& -e^{y/2} \exp \left(i\frac{me^y}{2}(|z|-z) \right)\; , \\
\psi^-(\tau=0,z) &=& e^{-y/2}
\left[ \exp \left( {i\frac{me^y}{2}\left(|z|-z\right) }\right) -1 \right]\; ,
\end{eqnarray}
and for the other one:
\begin{eqnarray}
\psi^+(\tau=0,z) &=& e^{y/2}  \left[ 1- 
\exp\left(i\frac{me^{-y}}{2}(|z|+z)\right) \right]\; , \\
\psi^-(\tau=0,z) &=& e^{-y/2}  \left[
\exp\left(i \frac{me^{-y}}{2}(|z|+z) \right)
\right]\; .
\end{eqnarray}

Because the $z$-dependent coefficients in the equation would make any
explicit scheme unstable, we discretise \eq \nr{eq:eomtauz}
implicitly\footnote{The
discretisation of a partial differential equation is said to be ``implicit''
when the time derivative of the unknown function at a given timestep
 depends on the unknown function at the next timestep.}:
\begin{eqnarray}
\la{eq:discr}
&&
\frac{1}{2\ud\tau}
\left[
\psi^\pm(\tau+\ud\tau,z)-\psi^\pm(\tau-\ud\tau,z) 
\right]
\nonumber\\
&&\quad=
\mp \frac{\sqrt{\tau^2+z^2}\pm z}{4 \tau \ud z}\bigg[
\psi^\pm(\tau + \ud \tau,z+ \ud z)+ \psi^\pm(\tau - \ud \tau,z+ \ud z)
\nonumber\\
&&\qquad
-\psi^\pm(\tau + \ud \tau,z- \ud z)- \psi^\pm(\tau - \ud \tau,z- \ud z)
\bigg]
-i m  \frac{\sqrt{\tau^2+z^2}\pm z}{\tau} \psi^{\mp}(\tau,z)\; .
\end{eqnarray}

Now the $\pm$-components are stored at different timesteps:
$\psi^+(\tau); \ \psi^-(\tau+\ud \tau)$. This saves memory
compared to storing the spinor at two timesteps,
while the discretisation is still  second order accurate in $\ud \tau$.
It seems that it is critical for the stability of the algorithm to also discretise
the endpoints to second order accuracy in $\ud z$.
Thus, when in \eq \nr{eq:discr} we have used the centered
difference
\begin{equation}
f'(z) \approx \frac{1}{2\,\ud z} \left[ f(z+\ud z) -f(z-\ud z)\right],
\end{equation}
for the points inside the lattice,
for the edges of the lattice \eq \nr{eq:discr} must be modified to use a one-sided
second order accurate difference:
\begin{equation}
f'(z) \approx \frac{1}{\ud z} \left[ 2 f(z+\ud z) - \half f(z+2\ud z) -\frac{3}{2}
f(z) \right].
\end{equation}
This prescription could be described as a free boundary condition.
Note that it would be quite unphysical to impose periodic boundary
conditions in the $z$-direction. Technically this shows up e.g. in the
fact that the coefficient in front of the $\partial_z$-term in
\eq\nr{eq:eomtauz} would be discontinuous at such a periodic boundary.

\eq\nr{eq:discr} forms a system of linear equations for
$\psi^\pm(\tau+ \ud \tau)$ in terms of the known values
$\psi^\pm(\tau-\ud \tau)$ and $\psi^\mp(\tau).$ The system is almost
tridiagonal and can be efficiently solved using LU-decomposition, 
leading to an algorithm which is slower than the corresponding explicit
discretisation only by a constant factor.

After having solved numerically the Dirac equation to find the spinor
at some finite proper time $\tau$, we must project out the positive
energy part of the wavefunction to find the amplitude for production
of a pair at rapidities $y,y'$. This is given by
\begin{equation}\la{eq:projsum}
M(y',y) = \tau \int \frac{\ud z}{\sqrt{\tau^2 + z^2}}
\overline{u}(y') \exp \left( im\left[ \sqrt{\tau^2 + z^2} \cosh y'
- z \sinh y' \right] \right) \gamma^\tau
\psi(\tau,z),
\end{equation}
where
\begin{equation}
\gamma^\tau = \gt e^{-\eta \gt \gz}
= \gt \left(\cosh \eta - \gt \gz \sinh \eta \right)
= \gt \left(\frac{\sqrt{\tau^2+z^2}}{\tau} - \gt \gz \frac{z}{\tau} \right)
\end{equation}
and $\tau /\sqrt{\tau^2 + z^2}$ is the Jacobian.

\begin{figure}[!htb]
\begin{center}
\includegraphics[width=0.7\textwidth]{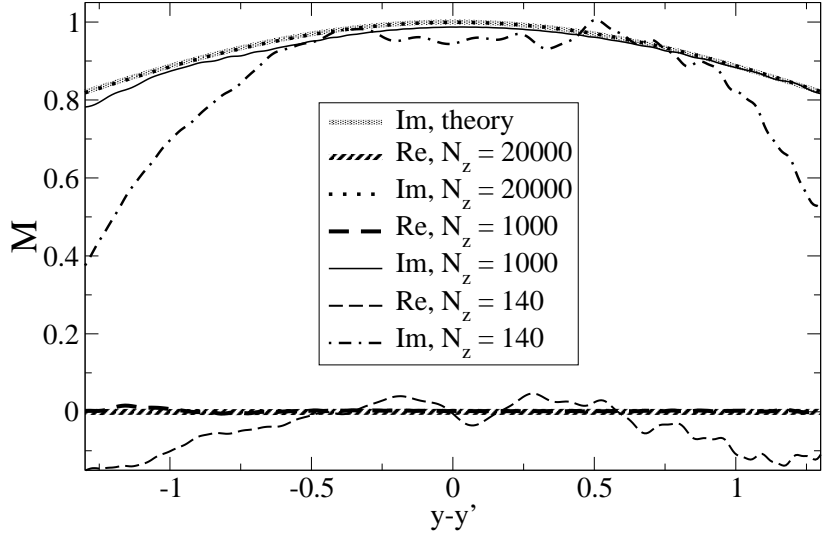}
\end{center}
\caption{\small Numerically calculated real and imaginary parts of the
free amplitude $M$ for one branch shown
for different lattice sizes in the $z$-direction.
Also shown is the analytical value $1/\cosh(y-y')$ of the imaginary part. The
analytical value of the real part is zero.}\label{fig:freeampli}
\end{figure}

\pagebreak[2]

In practice the integral, or sum in the discrete case, is an
oscillatory function for large $z$. In an analytical integration these
oscillations average to zero, but in a numerical calculation with a
finite extent in the $z$-direction this is harder to achieve. We have
used two different techniques to treat this problem. One is to
calculate the integral \nr{eq:projsum} for different upper and lower
limits $\pm z_\mathrm{max},$ and then take an average of the values
thus obtained over a range in $z_\mathrm{max}$ that contains several
periods of oscillation.  The other technique is to use wave packets
which are localised in the $z$-direction to slightly less than the
extent of the system in the $z$-direction.  This introduces an
uncertainty to the momentum in the $z$-direction or equivalently the
rapidity $y$, but the uncertainty is of the same order of magnitude as
the infrared cutoff from the size of the $z$-lattice.

The range of rapidity $y$ that can be reached in this coordinate
system is limited by two things. The finite lattice spacing in the
$z$-direction gives an ultraviolet cutoff for $p^z,$ implying that we
must have $\sinh y \lesssim {1}/({m \ud z})$.  Our method also
requires that the values of $\eta$ of the order of the momentum space
rapidities studied be covered by the finite lattice in the
$z$-direction.  This translates into the requirement $\sinh y \lesssim
{z_\mathrm{max}}/{\tau}.$ \fig\ref{fig:freeampli} gives an idea of how
close our results are to the analytically known result (for one
branch) $M(y',y) = i/\cosh\fr12(y-y')$ (see Appendix \ref{sec:ampli}).

%
\section{Dirac equation in 1+1 dimensions; external field}
\la{1+1external}

First let us note that this is not a calculation of pair production
from the Weizs\"acker-Williams fields of two currents on the light
cone, because such a thing does not exist in two spacetime dimensions.
Assuming an external current
$$
J^+=e_1\delta(x^-),\quad J^-=e_2\delta(x^+),
$$
one can namely solve $\partial_\mu F^{\mu\nu}=J^\nu$ for the only
component $E=F^{+-}$ to be $E=-e_1\Theta(x^-)+e_2\Theta(x^+)$. There
is just a constant electric field off the light cone.  Instead, our
purpose is to impose by hand an external field to test our numerical method
and to model the projection of the real 3+1 dimensional physics on the
longitudinal dimension. The parameters of the calculation will be $m$,
the mass parameter of the Dirac equation, $Q_s$, another mass
parameter describing the proper time variation of the external field
and $c$, a dimensionless parameter describing the strength of the
external field. To effectively describe the omitted transverse
momentum effects one might take $m$ to be not very different from
$\qs$. For $c\ll1$ one is in the weak field domain and analytic
results can be obtained.

\begin{figure}[!htb]
\begin{center}
\noindent
\includegraphics[width=0.7\textwidth]{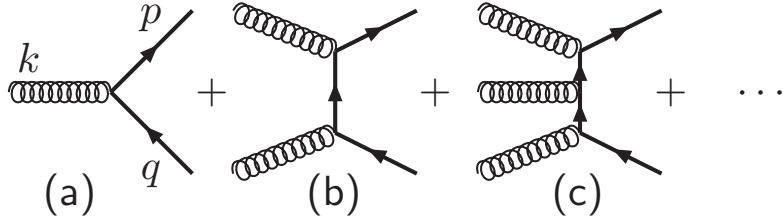}
\end{center}
\caption{\small
Diagrams contributing to the amplitude in the 1+1-dimensional toy model.
} \label{fig:diag}
\end{figure}

With the gauge choice $A_\tau=0$ and an external gauge field
$A_\eta(\tau)$ the Dirac equation \nr{eq:eomtauz} becomes:
\begin{equation}\label{eq:eomtauzphi}
\dtau \psi^\pm = \frac{\sqrt{\tau^2+z^2}\pm z}{\tau}
\left( \mp \partial_z \psi^\pm
- im \psi^\mp \right) \mp i \frac{gA_\eta(\tau)}{\tau} \psi^\pm.
\end{equation}
We shall study this for various choices of $A_\eta(\tau)$. The first choice,
motivated by the perturbative solution of the Yang-Mills equations\cite{kmw}
is
\begin{equation}\la{eq:extfield}
gA_\eta(\tau) = c \qs \tau J_1(\qs \tau).
\end{equation}
This form is special in that the Fourier transformed fields
corresponding to \nr{eq:extfield} can be given analytically:
\begin{equation}
gA^\pm(k^+,k^-)=\pm c \qs^2  \frac{i}{k^\mp+i\epsilon}\;
\frac{1}{2k^+k^- -\qs^2+i\epsilon( k^+ + k^-) }.
\la{aexplicit}
\end{equation}
Note that the pole structure is dictated by the requirement that
$A^\mu \sim \theta(x^+)\theta(x^-).$
Using this explicit form for the field in \nr{eq:extfield}
and the spinors in \nr{eq:planew2} one can write the
lowest order perturbative result (corresponding to diagram (a)
in \fig\ref{fig:diag}) for the amplitude:
\begin{equation}
M = - i g\overline{u}(p) \Asl(p+q) v(q)
\end{equation}
in the form
\begin{equation}\la{eq:resonans2}
M(\Delta y \equiv y-y') =  \frac{2 c \qs^2}{\cosh(\fr12 \Delta y)\;
[2m^2(1+\cosh\Delta y) -\qs^2 + i \epsilon]}
\equiv \frac{2 c \qs^2}{\cosh(\fr12 \Delta y)\;
[\hat s -\qs^2 + i \epsilon]}.
\end{equation}

\begin{figure}[!htb]
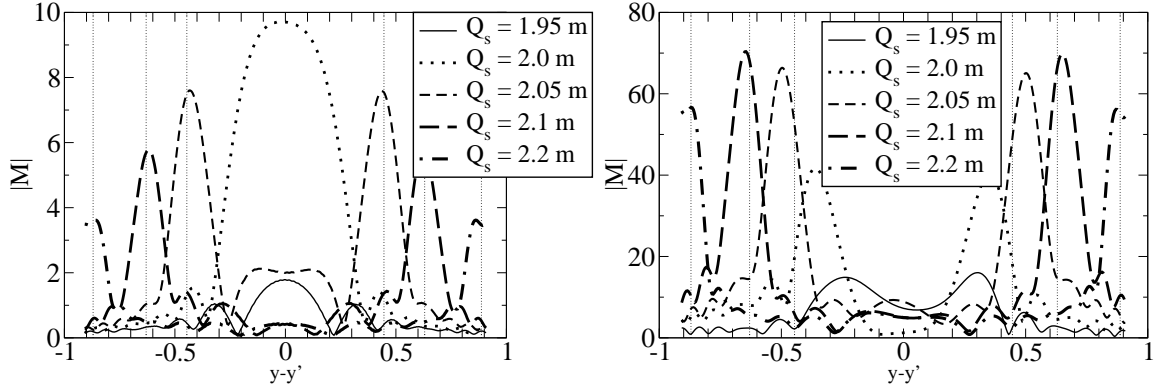

\begin{center}
\noindent
\includegraphics[height=145pt]{pertpeak.eps}
\includegraphics[height=145pt]{strongfieldpeak.eps}
\end{center}
\caption{\small Absolute value of the quark pair production  
  amplitude for different values of the oscillation scale $\qs$. Left:
  weak fields ($c \qs = 0.05 m$),
  the peaks are at the location given by \eq(\protect\ref{qscond}).
  Right: strong fields ($c \qs = m$) with the same values of $\qs.$ Peaks near
  the ``threshold'' $\qs=2m$ are shifted.  } \label{fig:ampli}
\end{figure}

\begin{figure}[!htb]
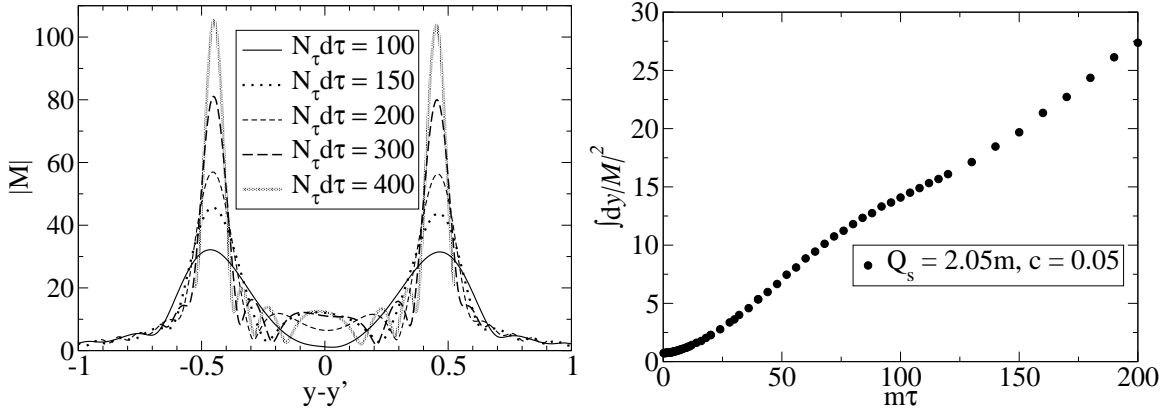

\begin{center}
\noindent
\includegraphics[width=0.49\textwidth]{tausize.eps}
\includegraphics[width=0.49\textwidth]{superthres.eps}
\end{center}
\caption{\small
Left: Absolute value of the amplitude for $\qs = 2.05 m$ and $c\qs = 0.5m$
when the projection is made at different
physical times $N_\tau d\tau$ at a fixed $d\tau = 0.05/m.$
Right: The number of produced pairs per unit rapidity,
$\int \ud \Delta y |M|^2,$ as a function of time. Note that  $\qs > 2m,$
so that the delta-function peak in \eq(\protect\ref{eq:resonans2}) dominates.
}\label{fig:tausize}
\end{figure}

Numerical results for the field \nr{eq:extfield} are shown in \fig
\ref{fig:ampli} at a fixed large time $\tau=N_\tau d\tau$, the time
dependence is studied in \fig \ref{fig:tausize}. In \fig
\ref{fig:ampli} the left panel shows the amplitude for weak fields and
the right one for strong fields.  For weak fields, $c \ll 1,$ one
expects the numerical result to coincide with first order perturbation
theory, as given by \nr{eq:resonans2}, which has a peak at $\hat
s=\qs^2$ or at
\begin{equation}
\cosh{\frac{\Delta y}{2}} = \frac{\qs}{2m}.
\la{qscond}
\end{equation}
For $\qs<2m$ this equation has no solution for $\Delta y$ and, in
fact, the numerical result is very small. Precisely at ``threshold'',
$\qs=2m$, there is a very strong single peak at $\Delta y=0$, the
quark and antiquark emerge at rest relative to each other. For
$\qs>2m$ there are two peaks corresponding to the two signs of
solutions of \nr{qscond}. These are well reproduced by the numerical
calculation. Due to the finite time the peak is not a delta function,
but is broadened. Physically, the amplitude peaks at pair invariant
mass = $\qs$ and, in 1+1d, the only way to give the pair this
invariant mass is to separate them in rapidity. In 3+1d the situation
is quite different, since then
\begin{equation}
\hat s\equiv (p+q)^2=2\omega_p\omega_q\cosh\Delta y + 2m^2-2 \pt \cdot \qt
\end{equation}
and the pair invariant mass can even be dominated by transverse momenta.

For stronger fields the numerical calculation sums over all orders in
the external field (all diagrams in \fig\ref{fig:diag}) and one does
not necessarily expect any peak structure. However, peaks still appear
(right panel of \fig \ref{fig:ampli}), although the location of the
peaks is shifted, especially near $\qs=2m$.

\fig\ref{fig:tausize} shows the dependence on the physical time
$\tau=N_\tau \ud\tau$, at which the projection to the final state is
done. One sees that the height of the peaks increases essentially
linearly in time, $M_\rmi{peak}\sim N_\tau$, while their widths shrink
somewhat.  The area of each of the peaks, $\int \ud y |M|,$ is
approximately independent of $N_\tau,$ showing that the resulting
rapidity distribution at asymptotic times contains two delta function
peaks in $\Delta y$.  The right panel shows that the integral $\int
\ud y |M|^2$, giving the number of produced pairs, grows approximately
linearly with $\tau.$ This monotonic increase is due to the fact that
the ansatz of \eq\nr{eq:extfield} behaves like $\sim\sqrt\tau$ at large $\tau$.

As a second example we shall consider a non-oscillatory and
exponentially decaying field 
\be \label{eq:expfield} 
gA_\eta(\tau)=c \qs \tau e^{-\qs \tau}.  
\ee 
This is actually simply reproduced from
the first ansatz of \eq\nr{eq:extfield} by taking a superposition of Bessel
functions $\omega \tau J_1(\omega \tau)$ with different frequencies
$\omega,$ which ``washes out'' the peaks at $\hat s=\omega^2$ in
\nr{eq:resonans2}. The appropriate weight factor is given by the
relation
\begin{equation}\la{eq:expdecay}
gA_\eta(\tau) = c \int_0^\infty \ud \omega
\frac{\qs \omega}{\left(\omega^2+\qs^2\right)^{3/2}}  \omega \tau J_1(\omega \tau)
= c \qs \tau e^{-\qs \tau}.
\end{equation}
Integrating over the matrix element \nr{eq:resonans2} (with
$\qs\to\omega$) one finds the perturbative matrix element
\begin{equation}\la{eq:expdecayampli}
M(\Delta y) = -\frac{ c \qs \shat}{\cosh (\half \Delta y)(\shat+\qs^2)^{3/2}}
\left\{
i\pi+2
\left[\frac{\qs^2}{\hat s}\sqrt{1+\frac{\shat}{\qs^2}}+
\ln \left(\sqrt{1+\frac{\qs^2}{\hat s}}+\sqrt{\frac{\qs^2}{{\shat}}}\right)\right]
\right\}
\end{equation}
Now there is no peak at $\hat s=\qs^2$ and, when plotted as a function
of $\Delta y$ for various $\qs/m$, $M(\Delta y)$ is a Gaussian-like
curve centered around $\Delta y=0$.

\begin{figure}[!htb]
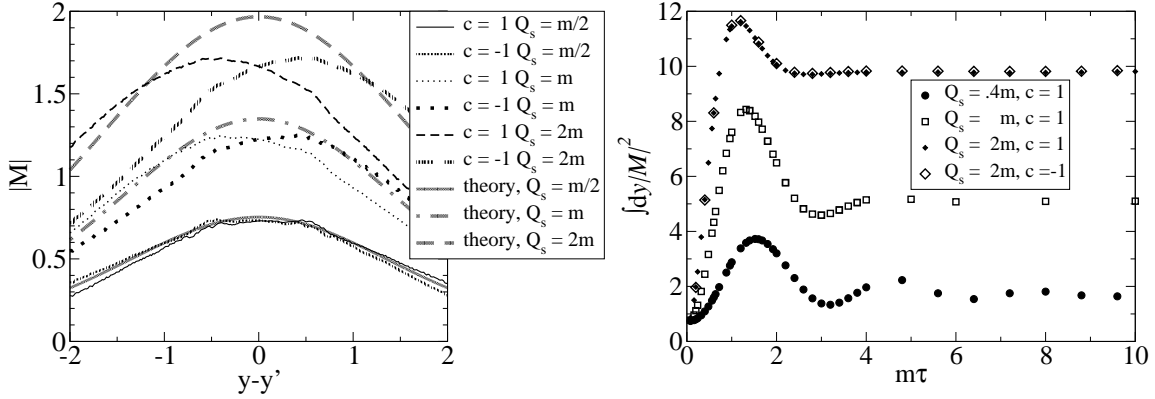

\begin{center}
\includegraphics[height=145pt]{expamplith.eps}
\includegraphics[height=145pt]{exptimedep.eps}
\end{center}
\caption{\small Left: Numerically computed amplitude for the 
  strong ($c=\pm1$) exponentially decaying field in
  \eq\nr{eq:expdecay}. The curves labeled ``theory'' are the weak
  field analytical result from \eq(\protect\ref{eq:expdecayampli}). Changing
  $c\to-c$ reflects the curves around $\Delta y=0$.  Right: The number
  of produced pairs per unit rapidity, $\int \ud \Delta y |M|^2,$ as a
  function of time for the exponentially decaying field.  }\label{fig:expdecay}
\end{figure}

Numerical results for the amplitude for $\qs \sim m$ and a strong
field $c=\pm1$ are shown in the left panel of \fig\ref{fig:expdecay}.
The general form of the rapidity dependence and the normalisation
agree quite well with the weak field formula \nr{eq:expdecayampli},
but the numerical curves are not centered exactly around $\Delta y=0.$
This asymmetry under $\Delta y\to-\Delta y$ is simply due to the fact
that under parity $A_\eta\to-A_\eta$ and thus the ansatz
\nr{eq:expfield} (as well as any non-zero $A_\eta$) breaks parity.  To
check that this is a real physical effect the numerical calculations
have been performed with both $c=1$ and $c=-1$. In the Dirac equation
this corresponds to $z\to-z$ and $\Delta y\to-\Delta y$ and the curves
obtained agree with this.  The right panel of \fig\ref{fig:expdecay}
shows the number of pairs produced, $\int \ud \Delta y |M|^2$.
Initially it rises $\sim \qs\tau$, then after a few oscillations,
caused by the coupling of the two frequency scales $\qs$ and $m$,
reaches a constant multiplicity level.
$\qs$ being a timescale of damping, the oscillation frequency is given by $m$.
The larger $\qs/m$, the larger is the multiplicity.  
This constancy is due to the fact that the external field vanishes in a time 
$\sim 1/\qs$.

\section{Numerical tests}
\begin{figure}[!htb]
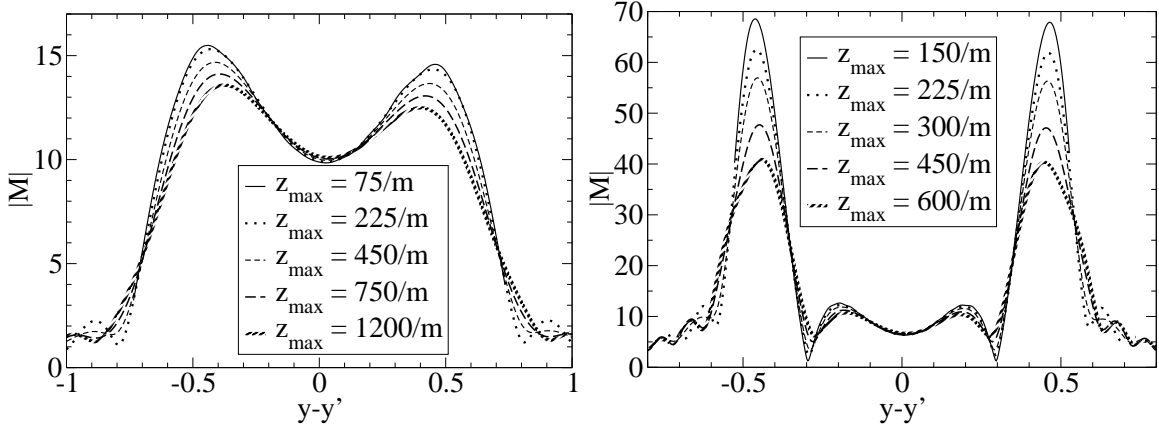

\begin{center}
\noindent
\includegraphics[width=0.49\textwidth]{zsize2.eps}
\includegraphics[width=0.49\textwidth]{zsize.eps}
\end{center}
\caption{\small Absolute value of the amplitude for different 
  sizes of the lattice size in the $z$-direction. The external field
  is a Bessel function \eq(\protect\ref{eq:extfield}) with a frequency slightly
  above the resonance condition, $\qs = 2.05 m.$ Left:
  $c \qs = 0.615m$ and $N_\tau \ud \tau = 60/m$. 
  Right:  $c \qs = 0.5m$ and $N_\tau \ud \tau = 100/m$. 
  There is some dependence
  on the lattice size, and the dependence is larger when the
  projection is done at a later time (right panel).
}\label{fig:zsize}
\end{figure}

Because of the delicate nature of the time-longitudinal dynamics, we
have performed various tests of the numerics.
In \fig\ref{fig:zsize} we study the effect of taking different
(physical) sizes for the lattice in the $z$-direction, or
$z_\textrm{max}$, using the Bessel function external field
\nr{eq:extfield}. Especially when the projection is done at a larger
time the amplitude depends somewhat on the size of the lattice.  The
difference in the multiplicity $\int \ud y |M|^2$ is, however, quite
small.  For the exponentially decaying field \nr{eq:expfield} the
lattice size effect is much smaller, almost unobservable on a plot
like \fig\ref{fig:zsize}. The effect of a finite $z_\textrm{max}$ on
the integral $\int \ud y |M|^2$ is larger for the exponentially
decaying field, because of the contribution from larger values of
$|\Delta y|$ that are unaccessible for a small $z_\textrm{max}$ (see
the discussion at the end of \se\ref{free1+1numerical}).

\begin{figure}[!htb]
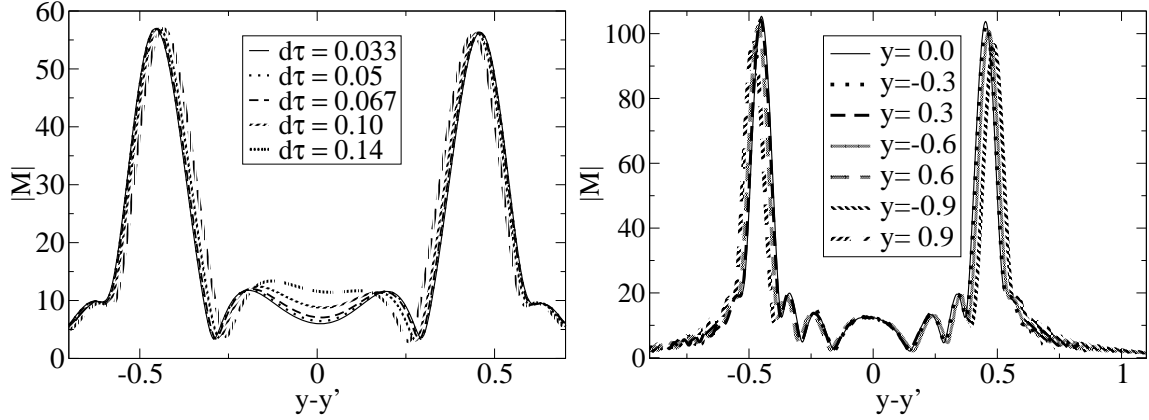

\begin{center}
\noindent
\includegraphics[width=0.49\textwidth]{taustep.eps}
\includegraphics[width=0.49\textwidth]{boostinv.eps}
\end{center}
\caption{\small
  Left: Absolute value of the amplitude
  using different timesteps at fixed physical size $N_\tau
  d\tau=400/m$.  Right: amplitude for different values of the antiquark 
  rapidity $y$;
  to check that the numerical calculation reproduces the boost
  invariance of the solution.  Both plots have an oscillating field 
  with $\qs = 2.05 m$ and $c\qs = 0.5m$ but different lattice sizes in
  the $z$-direction.}\label{fig:taustep}
\end{figure}

\begin{figure}[!htb]
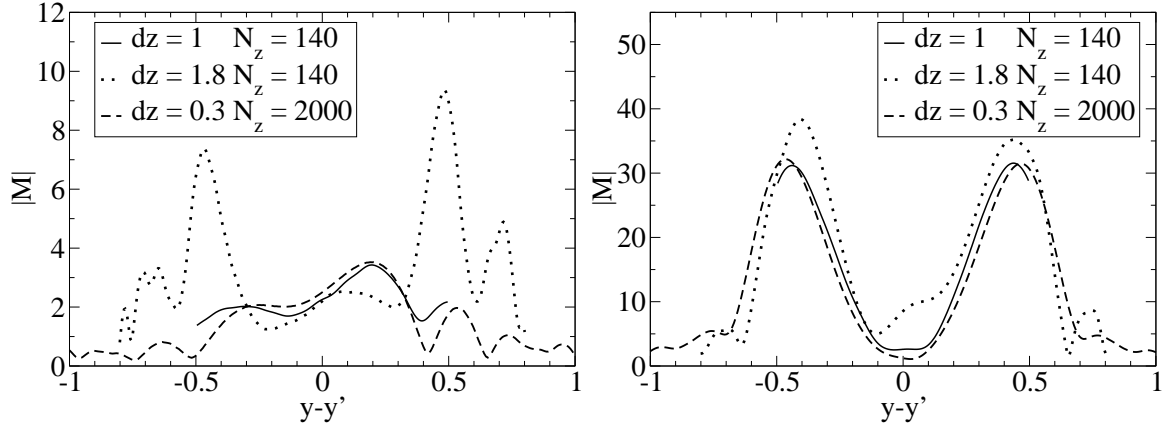

\begin{center}
\noindent
\includegraphics[width=0.49\textwidth]{smalllatperus.eps}
\includegraphics[width=0.49\textwidth]{smalllatpeak.eps}
\end{center}
\caption{\small Absolute value of the amplitude for smaller lattice sizes in
the $z$-direction. Left: $\qs$ below resonance ($\qs = m $ and 
$c\qs = m$). Right: $\qs$ above resonance ($\qs = 2.05 m $ and 
$c\qs = 0.5 m$). Values of $z$ in units of $[1/m].$} \label{fig:smalllat}
\end{figure}

The left panel of \fig\ref{fig:taustep} shows the dependence on the
timestep used.  In the right panel of \fig\ref{fig:taustep} we choose
different rapidities $y$ for the antiquark and compute the
distribution in the quark rapidity $y'$. The outcome is always a
function of $y-y'$, which shows that our numerical method preserves
the boost invariance of the result to a good accuracy.  In
\fig\ref{fig:smalllat} we explore the accuracy that can be reached
with lattices small enough to make a full 3+1-dimensional computation
realistic.  Although the computational requirements of a
3+1d-simulation are quite hard, we believe that with a careful choice
of discretisation parameters it is possible to extract some physical
results from a full 3+1d numerical calculation.

%
\section{Conclusions}
\la{conclusions} In this paper, we have set up the framework for
computing q\qbar pair production in ultrarelativistic heavy ion
collisions in the classical field model with an ensemble of quantum
initial conditions. This is an important theoretical problem,
especially in view of nonperturbative chemical thermalisation, for
which one needs a sufficient number of q\qbar pairs from the
dominantly gluonic initial state. However, the calculation is
technically complicated, involving the numerical solution of both the
gauge field equations of motion and the Dirac equation in the
background gauge field. Even the formulation of the initial condition
proves to be nontrivial since the natural variables $\tau,\eta$ cannot
in the limit $\tau\to0$ give a dimensionful longitudinal variable,
which one needs for longitudinal Fourier transforms.  In view of this,
we have in this paper limited ourselves to giving only the initial
condition for the full 3+1d problem but considered in numerical detail
only a 1+1d version of the model obtained after truncation of the
transverse dynamics. In this model both rapidity distributions and the
total number of produced pairs were computed for two forms of the
gluonic external field.

The work carried out here has solved the conceptually most complicated
part of the full 3+1d problem, the formulation of the initial
condition and the treatment of the longitudinal dimension together
with the proper time. The inclusion of transverse dynamics is
computationally demanding but otherwise straightforward. When that
part is completed, one will be in
a position to make  meaningful statements about the
nonperturbative production of q\qbar pairs in heavy ion collisions.

Future tasks include a full 3+1-dimensional treatment of also the
gauge field equations of motion and ultimately also including feedback
from the q\qbar sector to gluons, i.e., formulating and solving
coupled equations of motion. This will be a very challenging task.

%
\section*{Acknowledgements}

This work was partly supported by the Academy of Finland, Contract
no.\ 77744 and the European Community Integrated Infrastructure
Initiative Project "Study of Strongly Interacting Matter" Contract No
RII3-CT-2004-506078.
T.L. was supported by the Magnus Ehrnrooth Foundation and 
the Finnish Cultural Foundation.
We wish to thank J.P. Blaizot, D. Dietrich, E. Iancu, L. McLerran,
A. Peshier, K. Tuchin and R. Venugopalan for discussions on this and
closely related issues.
\appendix

\section{Dirac equation in curved coordinates}

Let us denote the flat coordinates $t,z$ or $x^0,x^3$ by Latin indices
$a,b,\ldots$ and the curved ones $\tau,\eta$ by Greek ones: $\mu,
\nu,\ldots$.  The flat metric is $\eta_{ab} = \textrm{diag}(1,-1)$ and
the curved one $g_{\mu\nu}= \textrm{diag}(1,-\tau^2)$, $g^{\mu\nu}=
\textrm{diag}(1,-1/\tau^2).$ The nonzero Christoffel symbols for the
$\tau, \eta$ coordinates are \cite{makhlin} $\Gamma^\tau_{\eta \eta} =
\tau$ and $\Gamma^\eta_{\tau \eta} = \Gamma^\eta_{\eta \tau} =
1/\tau$.

Given some representation for the usual $\gamma$-matrices in flat
space, $\gamma^a,$ one can express the $\gamma$-matrices in curved
coordinates as $\gamma^\mu = e^\mu_a \gamma^a.$ The \emph{zweibein} 
$e^\mu_a$ relates the flat metric to the curved one by
$g_{\mu \nu} = e_\mu^a e_\nu^b \eta_{ab}$ and, conversely, 
$\eta_{ab} = e^\mu_a e^\nu_b g_{\mu \nu}.$
There is no unique
choice for the $e^\mu_a,$ reflecting the fact that
there are different ways one can attach a flat tangent space to each
point in spacetime.  We have mostly used the natural intuitive choice
for the \emph{zweibein}, namely $e^a_\mu = \partial_\mu x^a.$ But in
the $\tau,\eta$-coordinate system there is also another natural
choice, namely to take $e^0_\tau = 1$, $e^3_\eta = \tau$, so that
$\gamma^\mu$ do not depend on the coordinates. To preserve the local
Lorentz invariance of the Dirac equation, one must introduce a
\emph{spin connection}\cite{mottola}:
\begin{equation}
\Gamma_\mu =  \frac{1}{8} [\gamma^a,\gamma^b] e_{\nu a}
( \partial_\mu e^\nu_b +  \Gamma^\nu_{\mu \sigma} e^\sigma_b).
\end{equation}
In this case the spin connection has only one nonvanishing component:
\begin{equation}
\Gamma_\eta = \half \gt \gz.
\end{equation}
The free Dirac equation
$\left[i \gamma^\mu (\partial_\mu + \Gamma_\mu)  - m \right] \widetilde{\psi} = 0$
in this case becomes:
\begin{equation}
\left[i \left( \gt \dtau + \frac{\gz}{\tau} \deta + \frac{\gt}{2\tau} \right)
 - m \right] \widetilde{\psi} = 0.
\end{equation}
Here we have introduced the spinor $\widetilde{\psi}$ defined with
this choice of the \emph{zweibein}. It is related to the usual flat
space spinor by $\widetilde{\psi} = e^{- \half \eta \gt \gz} \psi.$
The plane wave solutions \nr{eq:planew1} and \nr{eq:planew2} now have a
form that makes boost invariance manifest:
\begin{equation}
\widetilde{\psi}_{(\pm)} (x) =
 \sqrt{m}e^{\mp i m \tau \cosh (y-\eta )}
\binom{e^{\half(y-\eta)}}{ \pm e^{\half (\eta - y) } }.
\end{equation}

\section{Amplitude}\la{sec:ampli}

We found that the solution of the free Dirac equation for the left
branch in \fig\ref{fig:spacet} in the future light cone is (\eqs
\nr{eq:psipfree} and \nr{eq:psimfree})
\begin{eqnarray} \la{eq:psifree}
\psi^+ &= &  - \sqrt{m} e^{y/2} \sum_{n=0}^\infty (i e^{y-\eta})^n J_n(m\tau) \\
\psi^- & =& \sqrt{m} e^{-y/2} \sum_{n=1}^\infty (i e^{y-\eta})^n J_n(m\tau).
\end{eqnarray}
To project to a positive energy state with rapidity $y'$, we calculate the amplitude
\begin{equation}
M_1(y',y) = \tau \int \ud \eta\, \overline{u}(y') e^{i m \tau \cosh(\eta -y')}
\gamma^\tau \psi(\tau,\eta).
\end{equation}
Inserting the form \nr{eq:psifree} and using the standard integral
\begin{equation}
\int_{-\infty}^\infty \ud  \eta \,e^{i \zeta \cosh ( \eta - y) - \nu \eta}
= e^{-\nu y} i \pi e^{i \nu \pi /2} H^{(1)}_\nu(\zeta)
\end{equation}
and the Wronskian relation
\begin{equation}
H^{(1)}_{n}(m \tau)  J_{n+1}(m\tau)-  H^{(1)}_{n+1}(m \tau) J_n(m\tau)
= \frac{2 i }{\pi m \tau}
\end{equation}
one gets
\begin{equation}
M_1(y',y) = \frac{-i}{\cosh \frac{y-y'}{2}}.
\end{equation}
The other right branch has the initial condition
\begin{eqnarray} \la{eq:psifree2}
\psi^- &= &  \sqrt{m} e^{-y/2} \sum_{n=0}^\infty (i e^{\eta-y})^n J_n(m\tau) \\
\psi^+ & =& - \sqrt{m} e^{y/2} \sum_{n=1}^\infty (i e^{\eta-y})^n J_n(m\tau)
\end{eqnarray}
and leads to a contribution which exactly cancels that from the left branch:
\begin{equation}
M_2(y',y) = \frac{i}{\cosh \frac{y-y'}{2}}.
\end{equation}
These two contributions and the way they cancel are exactly the same that come from
evaluating, in 1+1 dimensions, the Feynman diagrams in \fig\ref{fig:tchan}.


\end{document}